# Non-equilibrium melting of colloidal crystals in confinement

Enrique Villanova-Vidal [1], Thomas Palberg [1,*], Hans Joachim Schöpe [1] and Hartmut Löwen [2]

[1] Institute of Physics, Johannes Gutenberg University, Staudinger Weg 7, D - 55128 Mainz, Germany

[2] Institute of Theoretical Physics II, Heinrich Heine Universität, Universitätsstrasse 1, D-40225 Düsseldorf, Germany

* corresponding author, email: Thomas.Palberg@uni-mainz.de

We report on a novel and flexible experiment to investigate the non-equilibrium melting behaviour of model crystals made from charged colloidal spheres. In a slit geometry polycrystalline material formed in a low salt region is driven by hydrostatic pressure up an evolving gradient in salt concentration and melts at large salt concentration. Depending on particle and initial salt concentration, driving velocity and the local salt concentration complex morphologic evolution is observed. Crystal-melt interface positions and the melting velocity are obtained quantitatively from time resolved Bragg- and polarization microscopic measurements. A simple theoretical model predicts the interface to first advance, then for balanced drift and melting velocities to become stationary at a salt concentration larger than the equilibrium melting concentration. It also describes the relaxation of the interface to its equilibrium position in a stationary gradient after stopping the drive in different manners. We further discuss the influence of the gradient strength on the resulting interface morphology and a shear induced morphologic transition from polycrystalline to oriented single crystalline material before melting.

1. Introduction

Consider a slab of ductile, low-melting material confined between chunks of larger rigidity and higher melting temperature. Subject this composite solid to a temperature gradient, which will partially melt the slab material and add a collinear pressure gradient. Depending on the sign of latter the slab material will either move down the temperature gradient and successively solidify or move up and successively melt. The former case is frequently encountered at die-casting or strip casting of alloys where intermediate semi solid slurries are formed [1], the latter is realized e.g. in zone melting, but also may occur in the perfusion of low melting rock material into other geological horizons or even in the process of magma intrusion into chambers from below [2]. Detailed knowledge about such phase transitions far from equilibrium conditions and at the same time influenced by the restricted geometry is still poor as such systems are notoriously inaccessible for experiments and computationally extremely demanding due to the multiscale character of the problem. In the present contribution we suggest to investigate them in a mesoscopic model system using charged colloidal crystals in confinement which are simultaneously subjected to an external drive and a gradient in salt concentration.

Colloids in general offer a unique opportunity to precisely tailor particle interactions with experimental means, thus switching from repulsive to attractive, from short to long range, from spherically symmetric to dipolar or directed interactions [3]. A direct consequence of this tunability are fascinating options to mimic atomic behaviour on a conveniently accessible mesoscopic scale. Hard sphere colloids in organic solvents form supercritical fluids, noble gas like, close packed crystals or fragile glasses [4] while hard spheres with added non adsorbing polymers serve as weakly attractive model systems forming stable liquids and various ways of phase separation [5]. Like-charged spheres in e.g. water form fluids, crystals or amorphous solids [6, 7] which display metal-like elastic behaviour [8]. Here the spheres take the role of atoms and the counter-ions correspond to the (valence-) electrons. Finally, oppositely charged particles display a huge variety of salt structures [9, 10]. Hence a wide range of material classes is represented in the phase behaviour of colloids and their equilibrium properties have been investigated in detail.

Recently, interest shifted to non equilibrium processes with special focus on phase transitions kinetics and the influence of external fields [5, 11, 12, 13]. By contrast to atomic systems, equilibrium melting is induced through a variation in interaction strength (e.g. via composition, concentration of particles and/or the amount of screening electrolyte) but not by an increase in temperature. The latter remains practically constant since the solvent acts as an effective heat sink. Still, the location of phase transitions is given by the ratio between thermal and interaction energy. Thus several authors have demonstrated the existence of equilibrium fluid-melt interfaces in stationary gradients of different nature, including electrolyte strength, particle concentration or solvent composition [14, 15]. Further, due to the soft matter nature of colloidal crystals (yield moduli of a few Pa only) application of external electric or magnetic fields [16, 17] or mechanical stress [18, 19, 20, 21] (both often in combination with confinement or the presence of rigid surfaces) are well suited to induce both melting and freezing or manipulate the systems morphology in during solidification. Thus also the issue of non-equilibrium phase transitions can be addressed in a very flexible way.

In the present paper we present the outline and preliminary results on a novel kind of experiment. With this we go beyond previous work in several points. First we study melting rather than crystallization. Here much less is known about the mechanisms involved to start and dominate the phase transition kinetics in an anisotropic environment. By contrast to the rich literature on the quantitative kinetics of crystal growth [5, 11, 12], even a qualitative parameterization of the melting kinetics is missing completely. Still open issues are e.g. the role of dislocations, grain boundaries and free surfaces [22, 23, 24, 25] favouring either randomly distributed global or local initiation of the melting process. Second, stationary gradients of interaction strength had been introduced in various ways [14, 15]. We here combine such a gradient with an externally pressure difference, and thus for the first time move the solid with respect to the gradient. And finally, previous work had focused on bulk situations. We here explicitly study the influence of the induced motion within the parallel wall confinement on the phase transition process.

The paper is organized as follows. We first shortly introduce the sample and the experimental set-up including a characterisation of the gradient formation and the pressure driven crystal motion. After giving an overview on the complex temporal evolution of the samples, we present exemplary quantitative measurements of the melting kinetics. A theoretical model is established in the next chapter, which is able to predict the interfacial velocity in the presence of a constant drive and after stopping the drive. We finally turn to a description and discussion of the peculiar morphological differences observed for different gradient strengths and the morphologic transition observed in strong gradients and in the presence of shear along a solid wall. We conclude with a short discussion of the scope and range of our approach.

## 2. Experimental

*2.1 Sample cell*

Experiments were performed in a commercial Microlife$^{(R)}$ cell [Hecht, Germany] sketched in Fig. 1. Two reservoirs are connected by a thin slit. The reservoirs contain approximately 1.5ml of suspension, the actual parallel plate measuring chamber spans 47mm in x direction, has a width y = 7.5mm and a height z = 500μm. One reservoir contains ca. 1ml of mixed bed ion exchange resin (IEX) [Amberlite, Rohm&Haas, France]. The cell is filled with suspension from a peristaltic conditioning circuit, where the desired number density of particles and the initial salt concentration were adjusted. Typically the suspension is deionized, but saturated with gaseous $CO_2$. The cell is then disconnected from the circuit and tightly sealed with screw caps. For technical reasons an air bubble remains under each lid.

The cell conveniently meets the two main requirements of our experiment: the possibility to maintain a stationary gradient in salt concentration along the cell and the possibility to generate a pressure difference between the two reservoirs causing a slow flow of the suspension through the cell.

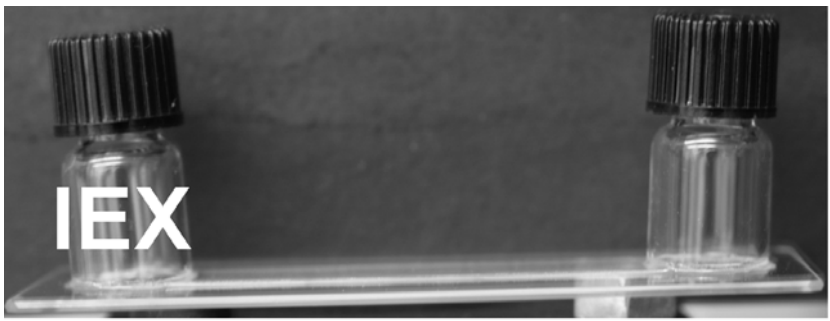

Fig. 1: Sample cell. The reservoirs contain approximately 3ml of suspension, the actual parallel plate measuring chamber spans 47mm in x direction, has a width y = 7.5mm and a height z = 500μm. In the filled state, one reservoir contains ca. 1ml of mixed bed ion exchange resin (IEX)

*2.2 Formation of a stationary salt gradient*

As only one of the reservoirs contains the IEX, the initially homogeneous salt concentration evolves into a gradient along the measuring chamber. The evolution of the gradient is not directly accessible. Therefore reference experiments were performed in a quartz cell of rectangular cross section (L×W×D = 150×10×1mm$^3$) connecting a reservoir filled with IEX with a second one without IEX. Here the local ion concentration was determined from either conductance measurements (performed with a set of 11 equispaced pairs of platinum electrodes placed at the small sides of the cell) or by photometry. We observed that the initial salt gradient was quite steep close to the IEX then extended and flattened. Linearity was typically reached within one to two weeks, slightly varying with initial salt concentration. This evolution was found to be independent of the type of salt chosen ($CO_2$, NaCl, Uranin). Using dissolved $CO_2$ as salt source has the advantage of keeping the reservoir salt concentration constant at a well defined level through the dissociation equilibrium $CO_2 + 2 H_2O \leftrightarrow HCO_3^- + H_3OH^+$. Over a length of 47mm the salt concentration drops linearly from c = $c_0$ = $1.1 \times 10^{-5}$ mol l$^{-1}$ in the IEX free reservoir with a $CO_2$ saturated state to c = $2 \times 10^{-7}$ mol l$^{-1}$ (from the self dissociation of water) at the IEX containing reservoir.

This gradient is stable until either all dissolved $CO_2$ has dissociated or the IEX becomes exhausted. Starting from $CO_2$ saturated conditions the former typically happens on a time scale of half a year, while with approx. 1ml of IEX the latter occurs within two to three months. Therefore the location of the melting salt concentration $c_M$ for a given particle number density is precisely determined for a time interval of say two weeks to two months. A second even stronger gradient is present above the IEX towards that reservoir´s gas bubble. The average salinity in this reservoir therefore is considerably smaller than in the other. Steeper gradients can be adjusted in the measuring chamber from elevated initial salt concentrations and will also pertain over weeks to months depending on the amount of added salt, the quantity of IEX present and the size of the reservoir.

*2.3 Generation of a pressure gradient*

Upon filling the sample is saturated with dissolved $CO_2$ but does not contain dissociated $CO_2$. Thus in a first process $CO_2$ dissociates leading to $c = c_0$. In the IEX free reservoir the dissociated $CO_2$ is slowly removed towards the IEX side across the cell. It is replenished by freshly dissociating $CO_2$ which in turn is replaced by gas from the air bubble in the lid. Hence there is a slight drop in the gas pressure on the IEX free side. On the IEX side an additional process is present. Due to the reduced average salinity the solubility of $CO_2$ drops [26] and neutral gas starts forming small bubbles adding to the gas bubble in the IEX side lid. Thus here a net pressure increase occurs. The situation is sketched in Fig. 2. To restore the pressure balance a very slow flow of suspension towards the IEX-free side occurs. We checked, that the motion is slow enough to not affect the salt gradient between the fixed positions of IEX and IEX free reservoir. Hence, the suspension is pushed up the salt gradient.

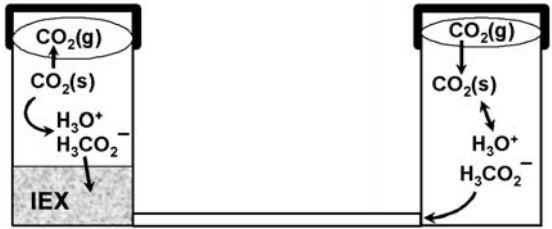

Fig. 2: Realisation of a pressure drive. One reservoir of the cell is filled with IEX. The IEX exchanges carbonate ions from dissociated $CO_2$ for hydroxyl-ions recombining with protons to water. At locally reduced salinity in the left reservoir the lowered solubility of $CO_2$ leads to degassing into the air bubble. At the opposite reservoir the salinity stays constant buffered by the reservoir of $CO_2$ in the air bubble, ready to dissolve and dissociate. A gas pressure difference is thus created exerting a force on the suspension, which drives it through the narrow part of the cell connecting the reservoirs.

The pressure difference cannot be measured directly. Rather its qualitative behaviour as a function of time was inferred from the induced motion of the sample through the flat part of the cell. We here exploited the effect that crystalline suspensions mostly showed a plug flow which allowed a convenient determination of the position of individual crystallites as a function of time from sequences of micrographs taken on the sample. Fig. 3a gives a graphic definition of measurable quantities. The positions $X_C(t)$ of individual crystals at similar x-position but differing y position can be monitored as a function of time. The positions $M(t)$ and $L(t)$ of the (curved) boundary between the two morphologies and of the (curved) crystal melt interface are determined as averages over the entire cell width, when only small curvatures are present. For strongly curved interfaces we favoured to first measure crystal or interface positions along the clearly visible stream lines (c.f. Fig. 4). The average extension of the WC region is obtained from $W = L - M$. Position measurements at individual times contain a comparably large uncertainty (typically ±100µm) due to the irregular fluctuations (presumably caused by coalescence of gas bubbles) superimposed on the continuous drift. Therefore, average propagation velocities were inferred from measurements performed over some hundred minutes.

By changing the initial $CO_2$ content, the size of the reservoirs, the amount of IEX and the size of the air bubbles, the pressure difference and the temporal evolution of the pressure difference and thus the drift velocity can be varied. For standard conditions (1ml IEX, $CO_2$ saturated suspension with no salt added) drift velocities of 0.5-5μm min$^{-1}$ were generated. Larger velocities could be obtained using an elevated initial salt concentration. Slower drifts are realized, if the IEX free side was left open to air. An example is shown in Fig. 3a, comparing the latter two cases. Typically, a roughly constant average drift velocity was obtained within one hour correlating with the time scale of deionization in the IEX reservoir. The drift typically pertained over a few weeks. It then slowly decreased to finally vanish when the degassing process, feeding the pressure difference, ceased.

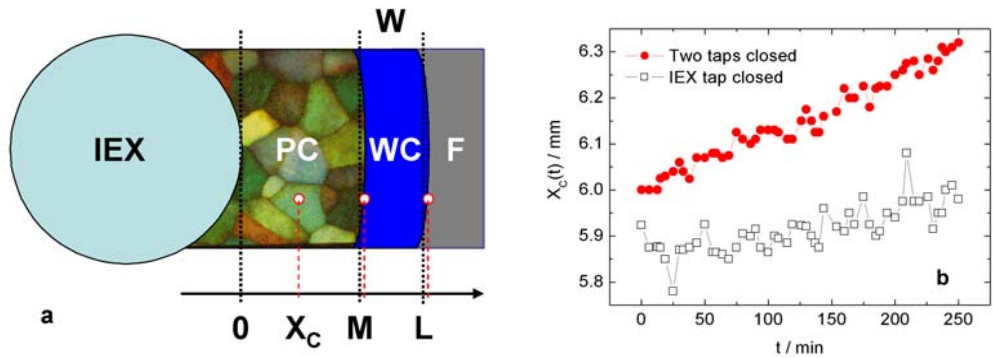

Fig. 3: a) definition of lengths to be measured in X direction: $X_C$: drifting crystal position; M: position of the morphological boundary between polycrystalline region (PC) and wall crystal region (WC); L: position of the crystal/melt interface. For flat interfaces y-averaged values were used (dotted black lines); for curved interfaces measurements were restricted to y-positions belonging to the same stream line (red dotted lines). b) Comparison of drifts as observed for a sample at $n = 12$μm$^{-3}$ under different conditions. Upper curve: initial salt concentration 20μM NaCl and both reservoirs tightly sealed; lower curve: no added salt, $CO_2$ saturation and the non-IEX side reservoir open to ambient air. Note the lower drift velocity and the increased fluctuations in the lower curve due to decreased pneumatic damping.

*2.4 Sample and initial sample conditioning*

The highly charged, small latex spheres (effective charge from elasticity measurements $Z_{eff}$ = 582±18, diameter from ultracentrifugation 2a = 122nm, standard deviation σ = 0.02) used in this work were synthesized by surfactant free emulsion polymerization and were a kind gift of BASF, Ludwigshafen. Samples were prepared from pre-cleaned stock suspensions filtered and stored over IEX. Samples were left at contact with air to obtain a thorough saturation with air-borne $CO_2$. For further conditioning they were then filled into a closed, peristaltically driven preparation cycle [27] connected to the measuring cell. The desired number densities were adjusted by dilution with doubly destilled, but $CO_2$ saturated water under control of static light scattering. Number densities $n$ were chosen to be above the equilibrium melting density at deionized conditions ($n_M(c = 0) \approx 0.1$ µm$^{-3}$). The suspension was then deionized again to remove the dissociated $CO_2$. During conditioning a meta-stable shear melt existed throughout the circuit and cell. The cell was then disconnected, and, if desired, a small amount of salt solution was mixed into the suspension. After introducing the IEX on one side, the cell and was sealed with air tight screw caps. This defines t = 0. For observation the cell was mounted on the stage of an inverted microscope (IRB, Leica, Wetzlar, Germany) and monitored with a high resolution CCD camera. Two observation modes were regularly applied: polarization and Bragg microscopy [12]. Data were stored in a computer and image processed to quantify sample structure, morphology and motion.

### 3 Melting experiments

*3.1 General observations*

We investigated samples with no added salt or with an initial salt concentration of 5-50µM NaCl. All samples readily crystallized within a few seconds after sealing and placing the cell on the microscope stage. Large $n$ samples formed polycrystalline solids, low $n$ samples large single crystals *via* heterogeneous nucleation at the container walls. For large $n$ and low salt initial concentrations these crystals were found to be stable. For low $n$ samples and large $n$ large salt samples, the crystallites melted again after a few ten minutes, showing a characteristic swiss cheese morphology indicative of a homogeneous release of ions during $CO_2$ dissociation. An example is shown in Fig. 6a. The melting time scale further compares well with the results of reference experiments on the conductance of pure water. There a homogeneous concentration of c = $c_0$ was typically reached within an hour.

Independent of the initial melting stage, the IEX immediately starts to remove salt next to the IEX reservoir. This locally leads to a steep gradient of salt concentration, which extends and flattens with time. In this gradient, starting next to the IEX the molten samples re-crystallized. The re-crystallized volume increased with time displaying a sharp boundary to the melt state. For combinations of $n$ and $c_0$ close to the equilibrium melting conditions the crystal melt boundary advanced though the complete cell. Only for combinations corresponding to an equilibrium fluid, the crystal-melt interface advanced to a position well accessible in the optical part. For $CO_2$ saturated initial conditions this was observed for $3µm^{-3} \leq n \leq 15µm^{-3}$, for elevated salt concentrations this occurred at correspondingly larger $n$.

The speed of the advancing interface is determined by both the drift velocity and the melting or growth velocity at the crystal-melt interface. This latter velocity is determined by the local salt concentration and hence coupled to the temporal development of the salt gradient. Also the shape of the advancing interface is determined this way and variations of either drift velocity or salt concentration in y-direction may lead to a curved interface. In practice, interfaces were frequently observed to be sharp and only show a mild symmetric variation by less than a two mm across the complete y direction. Other samples, however showed considerable curvature and differential advancement at different y-positions. In some of these we could discriminate bent rows of crystallites emerging from the IEX. Resembling the flow traces of glaciers, the crystal production and drift history is captured in such stream lines of crystals. On the other side in some cases the interface advancement is of different speed for different y-positions, while at the same time the drift velocity is stationary across the whole crystal. This indicates a variation of the salt concentration in y-direction. Also the observation, that even late stage equilibrated interfaces often remain curved, supports this suggestion. Typical examples of both effects is given in Fig. 4a-e for a suspension of $n$=12μm$^{-3}$ and an initial salt concentration of 20μM NaCl.

The overall temporal evolution in general is quite complex and differs in details for different initial conditions. Still, a standard scenario can be identified for the interface forming systems: After initial melting and formation of a crystal melt interface close to the IEX, the advancement of the interface continues for some 10 to 30 days, depending on the initial conditions. During this time the salt gradient is fully established and after some 15 to 30 days the drift gradually ceases. The interface position then either becomes stationary or shows a very slow retreat, to again stabilize after one or two weeks. The final position is found to be stable for an extended time, until the IEX becomes exhausted. For a typical volume of IEX of 1.5ml this takes about two to three months. After that the crystal melt interface recedes and the crystals disintegrates until the sample is completely molten. With our set-up we may observe both the melting of crystals pushed up the gradient, when the melting salt concentration is beyond the crystal melt interface, and the columnar growth of crystals, when the melting salt concentration is ahead the crystal melt interface. Hence the two main processes of zone melting are covered. In addition we may also access an equilibrated melt-crystal interface over several weeks.

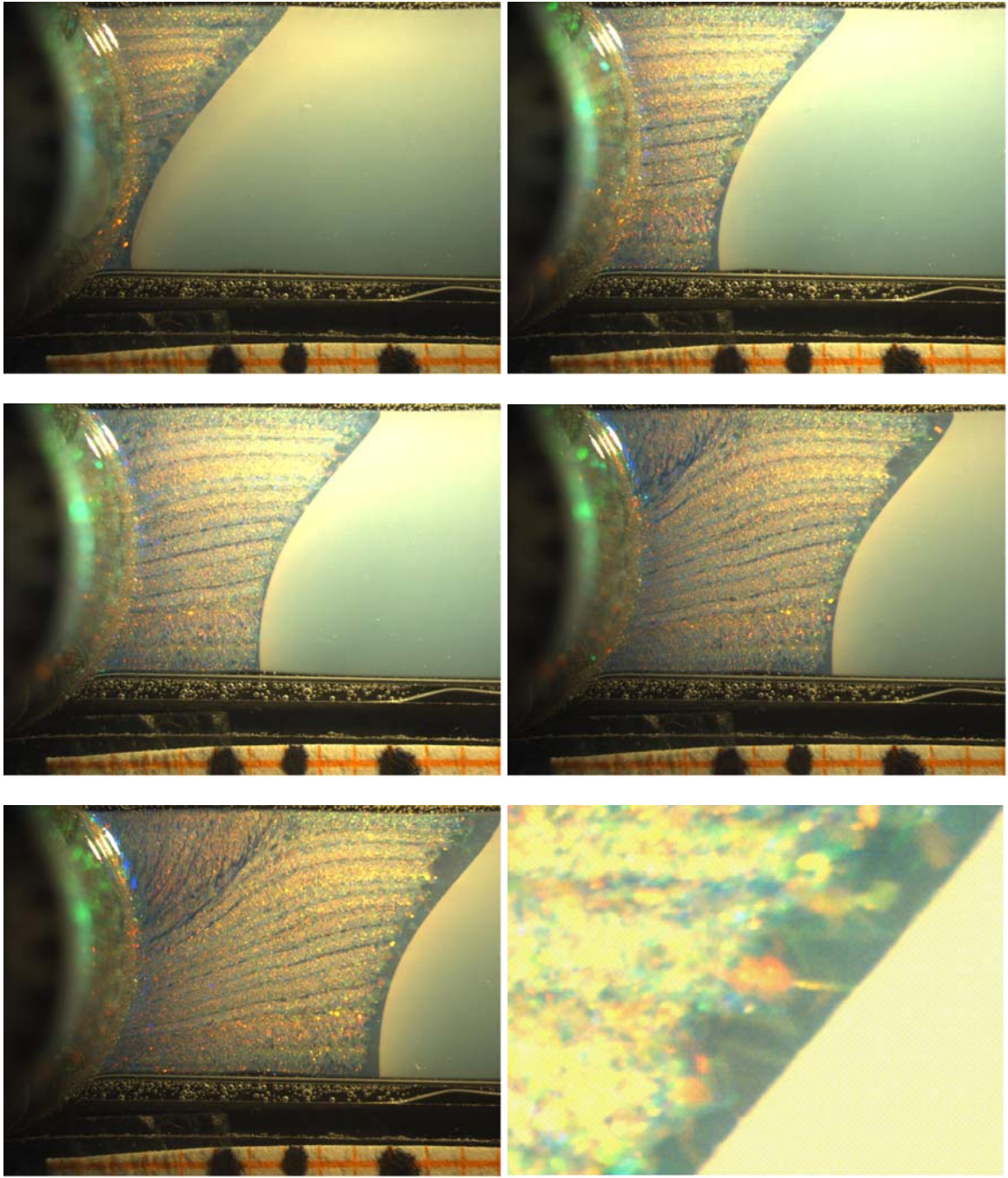

Fig. 4 a-e: Micrographs for a suspension at $n = 12\mu m^{-3}$ and initial salt concentration of 20μM NaCl for different times $t$ after filling. a) $t = 43h$, b) $t = 70h$, c) $t = 94h$, d) $t = 116h$ and e) $t = 139h$. Note the stream lines in the polycrystalline part of the advancing solid and the variation of the thickness of the darker, wall crystal region. f: magnification of the upper boundary region in c, comparing the smoothness of the crystal melt interface to the roughness of the morphologic transition boundary.

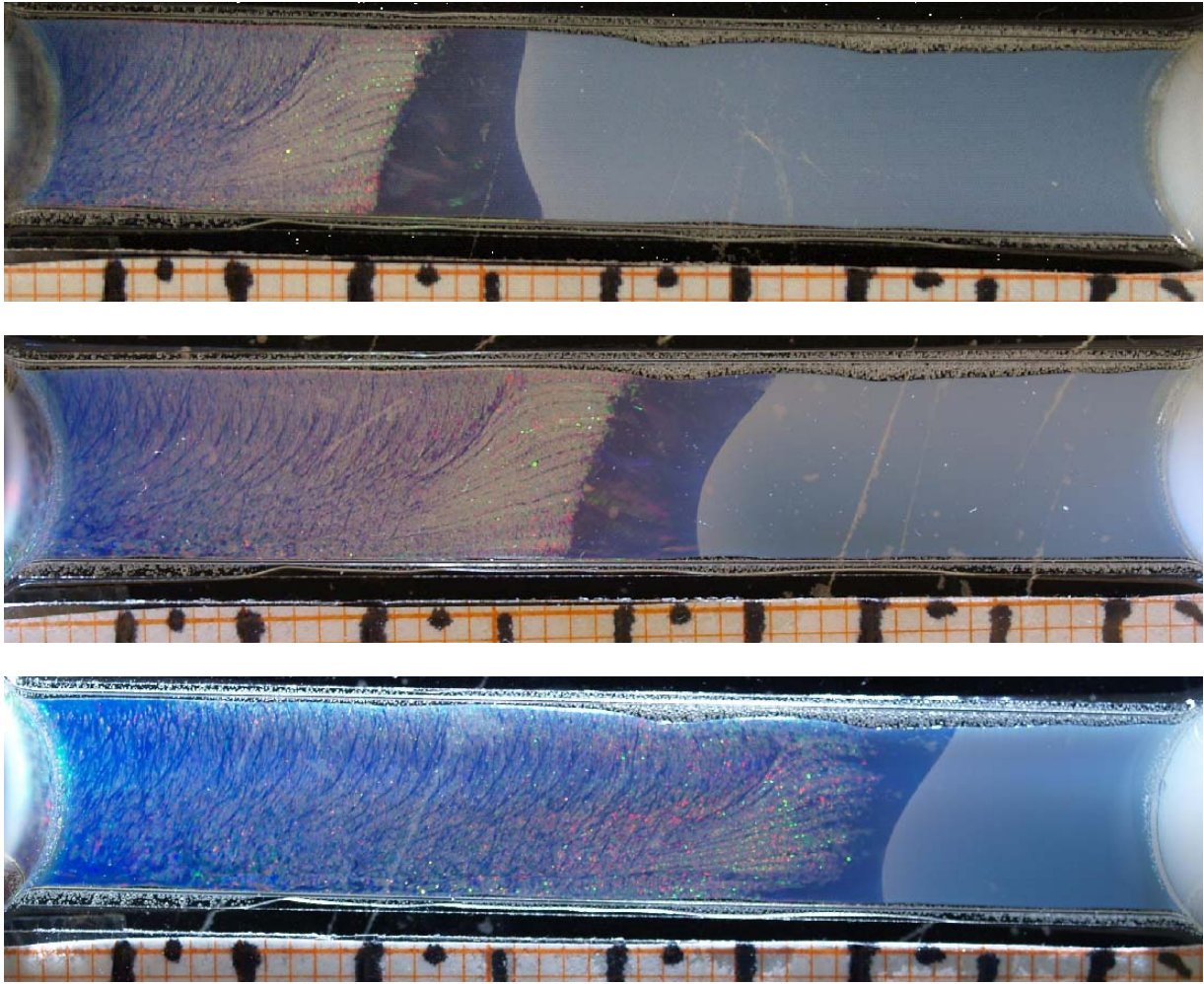

Fig. 5: As in Fig. 4 but now for times a) t = 12d, b) t 22d and c) t = 29d. Note the backward advancement of the morphologic boundary, presumably caused by vanishing of small crystals in the stream lines in favour of the large wall crystals.

*3.2 Melting kinetics*

In Fig. 5a-d we show close ups taken near the IEX-containing reservoir taken during the first five days. The IEX side is to the left. One may clearly distinguish the extensions of crystalline and fluid regions and in a qualitative way see their motion. From sample images the drifting crystal positions and the interface location were determined within stream lines with a temporal resolution of 10min. Results for $t$ = 50h are is shown in Fig. 6a-c for the interfacial position $L(t)$, the drifting crystal position $X_C(t)$ and their difference. From these measurements the velocity of crystal melt interfacial advancement $v_{front}$, the drift velocity $v_{drift}$ and the velocity $v$ inferred by least square linear fits. Obviously the crystal drift velocity is much larger than the interfacial velocity. Hence, $v$ = -45μm/h is negative and a *melting*-velocity. The crystal melts as it is pushed up the gradient.

Closer inspection of the linear fit with the data of Fig. 6b and c reveals that after two days the drift already slows down a bit as does the melting velocity, while the interface still advances quite linearly in time. The interfacial velocity slows noticeably several days later. This can be seen in Fig. 6d, where we show the interfacial position of our sample with a temporal resolution of one day over the entire measurement. For comparison we also include the data set for another sample at $n$= 6μm$^{-3}$ and no salt added. Here an indication of a relaxation behaviour is seen.

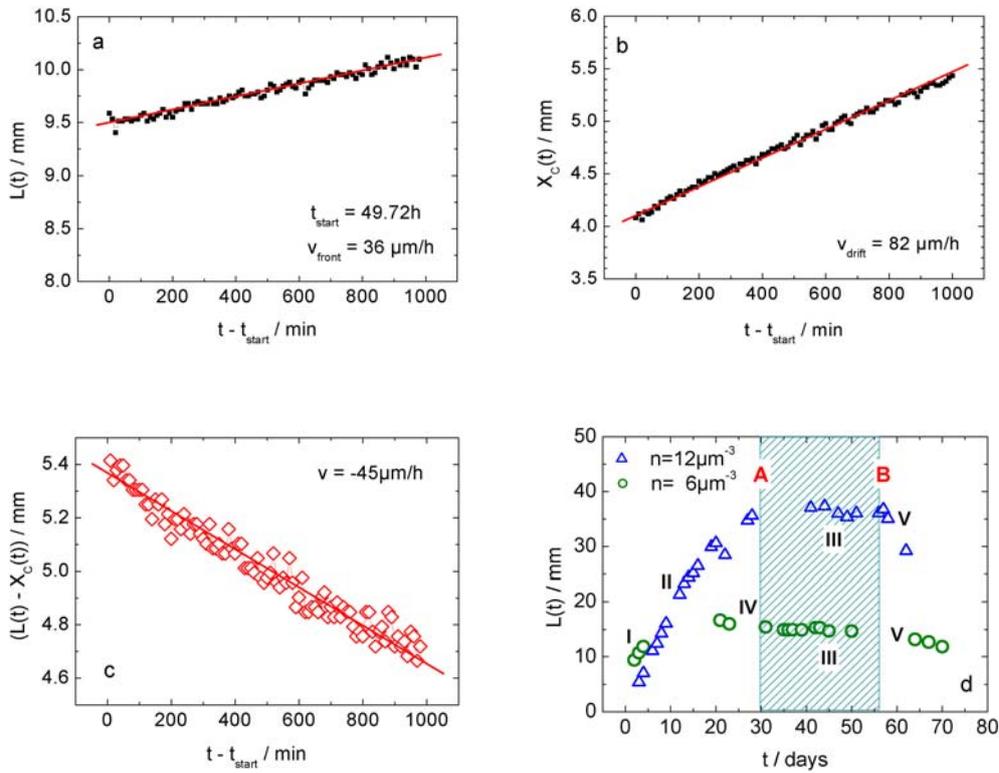

Fig. 6: Characteristic quantities for the sample at n = 12μm$^{-3}$ and an initial added salt concentration of 20μml$^{-1}$ shown in Fig. 4 and 5. All data were taken simultaneously in 10min. intervals for some 400min. a) Interfacial position L(t) b) drifting crystal positions X$_C$(t) and c) difference between these data. d) shows the extension of the solid region *L*(*t*) over the complete measurement for two samples of concentrations as indicated. Experiments on the sample of lower density were performed without added salt. The slowing approach of a stationary interfacial position is clearly visible. After presenting a stationary interface over more than a month (between A and B), the position recedes again due to exhaustion of the

IEX. Roman numerals mark the different stages of the temporal evolution. I: Initial melting; II advance of the interface; III stationary interface; V degradation. An additional relaxation stage (IV) seems to be visible in the data of the sample with smaller concentration. But this stage in general was found much less clearly distinguishable and reproducible.

**4 Theoretical model**

In a first approach to a theoretical description we make some simplifying assumptions. First we assume a fully developed stationary gradient, which is not perturbed by the advancing crystal. Second, we assume the drift velocity to be constant in time. And third, we focus on the case of balanced drift and melting velocities to stabilize the interfacial position. I.e. we investigate the case of an overheated crystal. All these assumptions may be released later on, e.g. by including the temporal development of the drift velocity and the gradient and allowing for crystal growth.

In order to establish a simple theoretical framework to classify and describe the zone melting behaviour. We assume a Wilson-Frenkel law [28, 29] for the modulus of the interface velocity v in terms of the dimensionless "overheating" $\Delta\mu / k_B T = (\mu-\mu_M) / k_B T$ [30]

$$v = v_\infty \left(1 - \exp(-\Delta\mu / k_B T)\right) \qquad (1)$$

Here, $v_\infty$ is the modulus of the limiting velocity at extreme overheatings, $\mu$ is the actual chemical potential of the colloids and $\mu_M$ denotes the colloid chemical potential at fluid-solid coexistence. The colloidal particles interact via a DLVO pair potential [31]

$$V(r) = V_0 \exp(-\kappa r)/r \qquad (2)$$

with $V_0 = Z_{eff}^2 \, k_B T \, \lambda_B$ and the screening parameter defined via $\kappa^2 = 4\pi\lambda_B(Z_{eff}\, n + 1000\, N_A\, c)$, r denoting the interparticle distance and $k_B T$ the thermal energy. $Z_{eff}$ is the colloidal effective charge number, and $\lambda_B = 7.8$Å is the Bjerrum length, $n$ denotes the colloidal number density, $c$ the molar salt concentration and $N_A$ is Avogadro's number. The DLVO potential provides a reasonable description of the interaction [32, 33] and the freezing and melting boundaries [7, 34, 35, 36] for charged stabilized colloidal suspensions. We then derive an approximate expression for $\Delta\mu / k_B T$ by mapping the Yukawa interaction potential onto a soft inverse power potential $\tilde{U}(r) = U_0 (\sigma / r)^4$. In this substitute potential, the dimensionless quantity $\Delta\mu / k_B T$ can only depend on differences in the dimensionless coupling strength $\Gamma = U_0 \sigma^4 \rho^{4/3} / k_B T$ such that $\Delta\mu / k_B T \approx \Delta\Gamma$. We map the potential $\tilde{U}(r)$ onto $V(r)$ by requiring that

$$\tilde{U}(d_{NN}) = V(d_{NN}). \tag{3}$$

where $d_{NN} = \sqrt{3} \times 2^{-2/3} \times n^{-1/3}$ is the nearest neighbour distance in the bcc crystal. Different salt concentrations lead to a variation of $\kappa$ and hence they map onto a different $\Gamma$ [37].

In a prescribed salt gradient $dc/dx$ the crystal-melt interface at position $L$ corresponds to a chemical potential difference of

$$\Delta\mu / k_B T = \alpha(L - L_0) \tag{4}$$

with $L_0$ denoting the interface position in equilibrium at zero velocity and

$$\alpha = 4\pi\lambda_B^2 Z_{eff}^2 \exp(-\kappa d_{NN}) \frac{dn_s}{dx} \frac{1}{\kappa} \tag{5}$$

Now we prescribe a time-dependent drift velocity $v_D(t)$. The actual interface velocity $dL/dt$ then is a difference of the imposed drift $v_D(t)$ and the melting interface velocity given by Eqn. (1)

$$dL/dt = v_D(t) - v_\infty(1 - \exp(-\alpha(L - L_0))) \tag{6}$$

which can be cast into the reduced form

$$dy/d\tilde{t} = g(\tilde{t}) - (1 - \exp(-y)) \tag{7}$$

with the reduced time $\tilde{t} = \alpha v_\infty t$, and $y(\tilde{t}) = \alpha(L(\tilde{t}/\alpha v_\infty) - L_0)$ and $g(\tilde{t}) = v_D(\tilde{t}/\alpha v_\infty)/v_\infty$. This differential equation can be solved for constant drift velocities $v_D$ and the solution is summarized as follows:

$$y(\tilde{t}) = \ln(\exp(A\tilde{t})(Ae^{y(0)} + 1) - 1)/A \tag{8}$$

where $A = v_D/v_\infty - 1$. For $v_D < v_\infty$ this implies that the growth saturates at $y(\tilde{t} \to \infty) = -\ln(1 - v_D/v_\infty)$ and this plateau value is reached exponentially in (reduced) time with a decay constant of $1/A$. For $v_D > v_\infty$, on the other hand, $y(\tilde{t})$ diverges linearly in time as $y(\tilde{t}) \approx (v_D/v_\infty - 1)\tilde{t}$. Finally, for the border case where the drift velocity coincides with the limiting velocity, $v_D = v_\infty$, we find an interesting logarithmic interfacial growth in time given by:

$$y(\tilde{t}) = \ln(\tilde{t} + e^{y(0)}) \tag{9}$$

For a general $v_D(t)$ there are only numerical solutions for the nonlinear differential equation (7). We have plotted the interface position as a function of time for different time-dependent drift velocities in Fig. 7. With the drift is switched on, the interface velocity first follows the drift but is then reduced by melting until the steady state is approached at balanced drift and melting velocity.

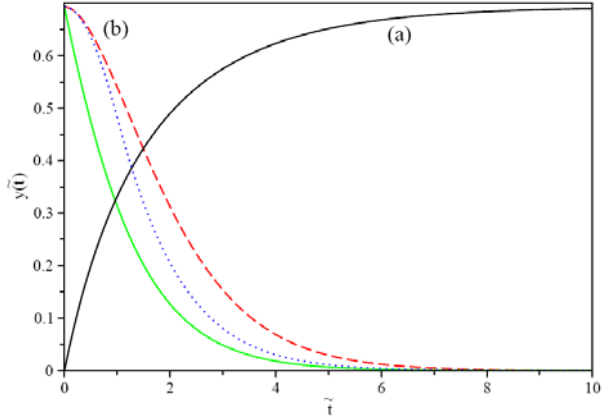

Fig. 7: Reduced interface position $y(\tilde{t})$ versus reduced time $\tilde{t}$ for various time-dependent drift velocities as obtained by a numerical solution of the nonlinear differential equation (7) for (a) y(0) = 0 and $g(\tilde{t}) = \Theta(\tilde{t})/2$; (b) y(0) = ln 2 after releasing g from g = 1/2 to zero as (i) $g(\tilde{t}) = \exp(-\tilde{t})/2$ (broken line), (ii) $g(\tilde{t}) = (1-\tilde{t})/2$ for $0 < \tilde{t} < 1$ and zero elsewhere (dotted line), (iii) $g(\tilde{t}) = \Theta(-\tilde{t})/2$ (full line). Here $\Theta(\tilde{t})$ denotes the unit step function.

At first sight, this behaviour appears rather similar to the evolution of the interface position in Fig. 6d. There are, however, some interesting differences. First, the experimental drift velocity is continuously reduced after two to three weeks. Hence the stationary interfacial position of the experiment (Stage III), is located closer than theoretically predicted or even at the equilibrium value. Second, at later stages there is a variation of the interface position by both melting and growth, which in addition varies with y-direction. This can be seen by closer inspection of Fig. 5 a and b. Focussing on the position of prominent features in the transition between the multicoloured polycrystalline region and the dark blue wall crystal region, one sees no variation in these features over a period of ten days. At the same time, however the lower part of the crystal melt interface has retreated, while the upper has advanced. Thus the wall crystal melts at some but grows at other y-positions, pointing at a variation of the salt concentration also across the cell. A y-direction variation of the salt gradient is difficult to implement without precise knowledge of the experimental boundary conditions. Here for the time being one has to restrict the quantitative comparison to cases of flat interfaces. Third, the presence of crystal growth in the experiments was not addressed here, but is covered by the model. Growth occurs, when in the symmetric Wilson-Frenkel law (Eqn. (1) chemical potential differences of opposite sign are used.

On the other side, the theoretically modelled situation of balanced drift and melting velocity is experimentally realized during the transition from stage I to stage II. During this period the drift is large and constant and the gradient is still steep. The interface first still recedes, but then for many hours the interface stays at approximately constant position and the microscopic analysis shows balanced velocities. Later, as the gradient evolves and the salt concentration at the interface decreases, the melting velocity has reduced below the drift velocity and the interface advances.

A first extension of the simple case discussed above was performed in a study of the relaxation from the stationary (velocity balanced) state to the equilibrium state after releasing the drift. Results are also shown in Fig. 7. Here three different cases corresponding to an instantaneous turn off of drift and to a linear and exponential decrease of drift are plotted. The relaxation time of the drift is mixing with the ultimate relaxation time $\alpha v_\infty$ of the interface. This curve in principle corresponds to the relaxation part of the $n = 6\mu m^{-3}$ sample shown in the experimental data of Figure 8. Unfortunately, at the present stage, the uncertainty in the experimental parameters is still too large to perform a detailed quantitative comparison by estimating the different time scales using e.g. eqn. (5) or to assess the logarithmic interface growth predicted for the border case in eqn. (9). This is left for future studies. We finally remark that close to equilibrium the nonlinear differential equation (7) can be linearized $(\exp(-y) \approx 1-y)$ and then the analytical solution becomes

$$y(\tilde{t}) = y(0)\exp(-\tilde{t}) + \int_0^{\tilde{t}} dx'(g(x')\exp(x'-\tilde{t})) \tag{10}$$

which can be used for an asymptotic study close to equilibrium if the drift is released.

To summarize the theoretical part: a simple theory assuming a Wilson-Frenkel law makes predictions for the time-dependent interface position in a salt gradient for a given constant drift velocity and can be used to study trends when other quantities are varied such as salt gradient, colloid effective charge and different time-dependent prescribed drift velocities. One further prediction is that - if the drift is released - the interface position approaches its equilibrium value exponentially in time. The associated decay time is $\alpha v_\infty$. Future extensions of the theory may conveniently be included. Future work e.g. will explicitly include the combined time dependence of drift and salt gradient to cover the experimental evolution of the interface position in more detail.

**5 Morphologic development.**

Two points concerning the sample morphology are of special interest. First, We observe rather different morphologies upon melting under different circumstances. Second, with a drift present, the polycrystalline material shows a morphologic transition before melting, which can be viewed as shear assisted alignment.

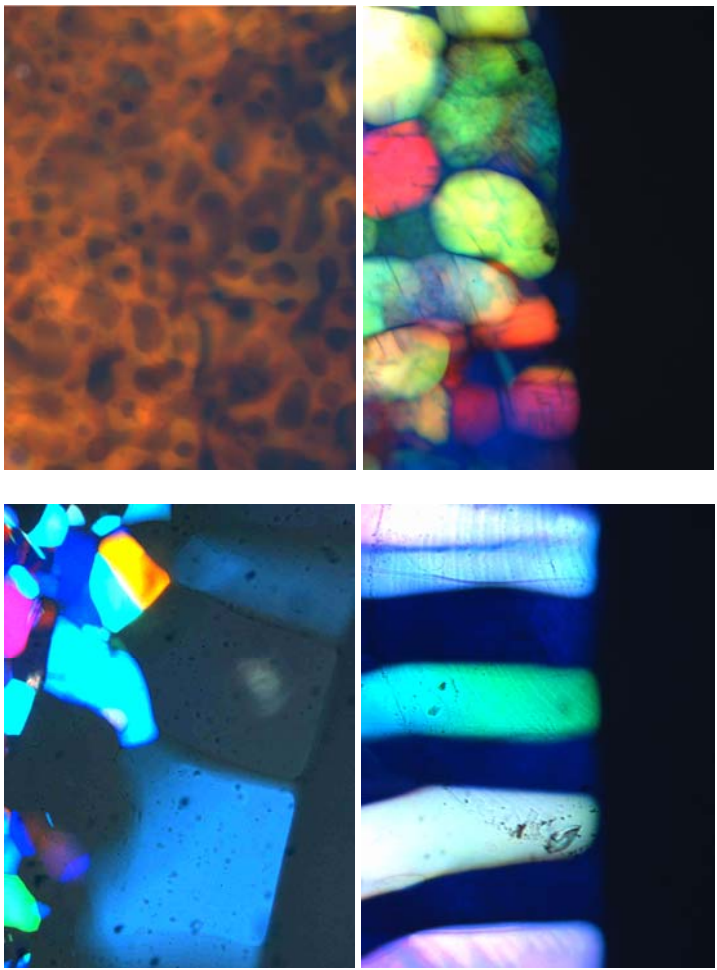

Fig. 8: different morphologies obtained by melting under different boundary conditions. a) swiss cheese morphology frequently< observed during initial melting by homogeneously released carbonate (image height ca. 8mm); b) polycrystals slowly advancing and melting in a shallow gradient resemble a beach of pebble stones. In an extended coexistence region, these melt inward. (image height ca. 4.6mm); polycrystals in a strong drift, approaching the crystal melt interface in a steep gradient. Before melting they transform to large wall crystals of like

crystallographic orientation. The wall crystals melt with straight sharp boundaries, showing kinks at their grain boundaries (image height ca. 5mm); d) columnar crystals grown in the absence of a drift (both taps open) and slowly melting in a moderate gradient as the carbonate concentration increases and showing extended kinks (image height ca. 5mm).

We first turn to the morphological differences observed under different conditions of melting. Concerning the initial melting stage, it is caused by a more or less homogeneous release of carbonate ions. Presumably for most of the cell no strong gradients in salt concentration are present. In this case we observe a peculiar swiss cheese like structure of the melting crystals. An example is given in Fig. 8a. Here white light illumination was used in combination with Bragg microscopy adjusted to meet the (111) reflection, leading to the characteristic rainbow like color distribution for different angles of incidence. Quite a different morphology is seen, when for low $n$ samples the polycrystalline material produced on the IEX side reaches the crystal-melt interface. Here the crystals melt starting from their grain boundaries, yielding a pebble like appearance of the sample shown in Fig. 8b obtained from polarization microscopy. At particle densities larger than $5.5\mu m^{-3}$ a wall crystal region appears close to the crystal melt interface. It is formed from the polycrystalline material. In this case of melting of oriented wall crystals in a gradient, a sharp interface is observed. Presence of several crystallites with quite parallel grain boundaries gives the crystal melt interface a characteristic teeth-like appearance (Fig. 8c). For most part of the crystal the interface is straight, but at the grain boundaries between different crystals one clearly observes the presence of kinks. Finally, in the absence of drift (both taps open) re-crystallization occurs by growth of columnar crystals, some of these also showing the characteristic colour of preferred orientation with the densest packed plane parallel to the cell wall, but others being oriented differently. The image in Fig. 8d gives an example for a low n case, where, like in Fig. 8b, the crystal melt interface is again more bent.

The main difference between the different situations is given by the steepness of the local salt gradient. Swiss cheese morphologies appear for vanishing gradients (c.f. Fig. 8a), while straight fronts appear for fast melting in a steep and straight gradient (Fig. 8c). Grain boundaries widening to pebblish morphologies with intact crystal interior as well as grooves entering at the grain boundaries mediate between these extremes. Here a moderate gradient is present which may initiates two effects: first salt may possibly enter the crystalline region better along the less ordered grain boundaries, second the grain boundaries may have a lower melting point than the well ordered crystals. Both gives rise to a realization of the coexistence region (between $c_F$ and $c_M$) with roundish crystals embedded in the coexisting fluid. The peculiar appearance as either pebbles or fingers relates to the initial crystal morphology after formation (polycrystalline or columnar). All these interface morphologies have been observed before: The pebblish appearance of polycrystals at coexistence has been noticed very early [38], columns were reported e.g. by Yamanaka et al. [14]. Even swiss cheese structures were observed in very carefully sealed, gradient free cells to evolve over several hours [25]. In addition recent studies using confocal microscopy showed that in nearly hard sphere crystals pre-melting occurs at defects, dislocations or grain boundaries providing a convenient way of starting the melt process [24]. The comparison made here, however, combines observations made on one species of charged spheres subjected to different kinds of gradients and thus allows to assess the solitary findings of previous work under a new common point of view.

Second, we have observed that after a few mm of drift the polycrystalline material transforms to rather uniformly oriented monolithic wall crystals of tooth-like to columnar morphology. In Fig. 5 and Fig. 8c the multicoloured polycrystalline region is clearly separated from the crystal melt interface by a dark blue region of wall based crystals. Also the boundary of the morphological transition is slightly bent and initially follows the shape of the crystal melt interface. The boundary is smooth on a large scale. But it is rough on a scale of several crystallites, as the transition often affects several neighbouring crystals at the same time or advances several crystals deep into the polycrystal following a stream line (c.f. Fig. 5). Concerning individual crystals, we have observed wall crystals emerging from coalescence of two or more polycrystals. We have seen individual crystals flip color as a whole but also sometimes seen a straight boundary to propagate through a crystal. The latter case may be some extreme case of coarsening simply enlarging the volume of the adjacent wall crystal and minimizing its surface. Coarsening is also present in the polycrystalline material as it is pushed up the gradient and enters regions of elevated salt concentration. There is, however, an important difference to normal bulk coarsening in a quiescent sample where random orientation of crystallites is retained. Due to the fluctuations of the gas pressure, in our case our coarsening suspension is sheared by a superposition of steady and oscillatory shear, with the largest shear rates at the walls. Additional scattering experiments reveal that beyond the zone of large polycrystals all wall crystals are oriented alike with the densest packed plane parallel to the confining walls and the densest packed direction parallel to the drift velocity. A similar orienting influence of an externally applied shear was found by computer simulations of sheared bilayers in [39, 40] and experimentally observed in re-crystallization experiments after directed shear melting [41] or even in continuous oscillatory shear [42]. Following these authors, also in our case the orientation of the dense packed planes is steered by the wall orientation, while further shear stress is minimization occurs by orienting the easy shear direction parallel to the drift direction. As we are working in a gradient of salt concentration, also the interaction strength decreases upon approaching the melting transition. Both shear modulus and yield modulus are directly related to the latter [43]. In the polycrystalline region the solid may still resist plastic deformation by shear and simply strains. At increased screening, however, it yields and is reoriented.

**6 Conclusions**

We have reported a novel versatile set-up to study non-equilibrium melting of charged sphere colloidal crystals either subjected to a spatially homogeneous increase in salt concentration or pushed up a gradient in salt concentration, such that a competition between drift velocity and thermodynamically controlled melting velocity occurs. In addition also equilibrated interfaces are obtained in a stationary salt gradient. We could demonstrate how kinetic measurements yield the involved velocities and establish a simple model to predict the advancement of the interface. While most observations are still in a preliminary state, we feel that the approach taken may in future significantly enrich our understanding of the melt process of colloidal model crystals. A number of interesting questions may be addressed. For example, small quantities of an additional species could be added to study the role of impurities in melting and their fate after re-solidification (crystal purification). Samples prepared close to the bcc-fcc-fluid triple point will complement previous studies on study phase selection during crystallization by others during melting with or without external fields applied. Also could the studies be extended to the melting of twinned bcc wall crystals obtained after directed shear melting at low *n*. From the technical point of view, one may replace the pressure drive by a piezomechanic drive to obtain a programmable drift and realize situations better comparable to theoretical predictions. Theory as well may be considerably extended to include a temporal variation of salt gradients and drift velocities. We therefore think to have presented a very flexible new concept to approach melting studies with model crystals. Moreover also equilibrated interfaces may be of great interest. In particular, an analysis of the radius of curvature of the equilibrated grooves of neighbouring crystals possibly offers an approach to fluid-crystal interfacial tensions, hardly accessible in other experiments.

In addition to charged sphere model crystals it will also be of interest to study crystals of other interactions, e.g. hard sphere crystals [4], ionic crystals of oppositely charged spheres [910] or crystals forming in attractive systems [5]. In mixtures, additional segregation phenomena may be expected. Also such systems should be equally well suited for studies of their melting kinetics and mechanisms in gradients, close to confining walls and under shear.

Any of these questions, however, is also of great interest to other scientific communities involved with non-equilibrium thermal processing of materials, e.g. metallurgy or geology. Here our approach bears the perspective of flexibly adjusting the boundary conditions to model closely related situations known from atomic matter. For instance the pressure drive mimics fluctuations inevitable e.g. in die or strip casting and presumably also in geologic processes. Further we have shown that the gradient characteristics directly influence the resulting melting morphology and that the simultaneous presence of shear and confining walls can significantly alter the melting process. Similar effects are also expected for atomic systems of spherically symmetric interactions. On the other side it would be interesting to study, whether the swiss cheese melting morphology can have a counter-part in atomic systems or is a colloid specific phenomenon. With our approach we hope to have stimulated fruitful discussions on the reported physical phenomena and demonstrated the potential of colloidal models to address non-equilibrium issues.

## Acknowledgements

We thank S. van Teeffelen and Erdal C. Oguz and Nina Lorenz for helpful discussions and BASF, Ludwigshafen for the kind gift of particles. We gratefully acknowledge financial support by the DFG (SPP 1120, SPP 1296, Pa459/12 and SFB TR6).

**Figure captions**

Fig. 1: Sample cell. The reservoirs contain approximately 3ml of suspension, the actual parallel plate measuring chamber spans 47mm in x direction, has a width y = 7.5mm and a height z = 500μm. In the filled state, one reservoir contains ca. 1ml of mixed bed ion exchange resin (IEX)

Fig. 2: Realisation of a pressure drive. One reservoir of the cell is filled with IEX. The IEX exchanges carbonate ions from dissociated $CO_2$ for hydroxyl-ions recombining with protons to water. At locally reduced salinity in the left reservoir the lowered solubility of $CO_2$ leads to degassing into the air bubble. At the opposite reservoir the salinity stays constant buffered by the reservoir of $CO_2$ in the air bubble, ready to dissolve and dissociate. A gas pressure difference is thus created exerting a force on the suspension, which drives it through the narrow part of the cell connecting the reservoirs.

Fig. 3: a) definition of lengths to be measured in X direction: $X_C$: drifting crystal position; M: position of the morphological boundary between polycrystalline region (PC) and wall crystal region (WC); L: position of the crystal/melt interface. For flat interfaces y-averaged values were used (dotted black lines); for curved interfaces measurements were restricted to y-positions belonging to the same stream line (red dotted lines). b) Comparison of drifts as observed for a sample at $n = 12\mu m^{-3}$ under different conditions. Upper curve: initial salt concentration 20μM NaCl and both reservoirs tightly sealed; lower curve: no added salt, $CO_2$ saturation and the non-IEX side reservoir open to ambient air. Note the lower drift velocity and the increased fluctuations in the lower curve due to decreased pneumatic damping.

Fig. 4 a-e: Micrographs for a suspension at $n = 12\mu m^{-3}$ and initial salt concentration of 20μM NaCl for different times $t$ after filling. a) $t = 43h$, b) $t = 70h$, c) $t = 94h$, d) $t = 116h$ and e) $t = 139h$. Note the stream lines in the polycrystalline part of the advancing solid and the variation of the thickness of the darker, wall crystal region. f: magnification of the upper boundary region in c, comparing the smoothness of the crystal melt interface to the roughness of the morphologic transition boundary.

Fig. 5: As in Fig. 4 but now for times a) t = 12d, b) t 22d and c) t = 29d. Note the backward advancement of the morphologic boundary, presumably caused by vanishing of small crystals in the stream lines in favour of the large wall crystals.

Fig. 6: Characteristic quantities for the sample at n = $12\mu m^{-3}$ and an initial added salt concentration of $20\mu ml^{-1}$ shown in Fig. 4 and 5. All data were taken simultaneously in 10min.

intervals for some 400min. a) Interfacial position L(t) b) drifting crystal positions $X_C(t)$ and c) difference between these data. d) shows the extension of the solid region $L(t)$ over the complete measurement for two samples of concentrations as indicated. Experiments on the sample of lower density were performed without added salt. The slowing approach of a stationary interfacial position is clearly visible. After presenting a stationary interface over more than a month (between A and B), the position recedes again due to exhaustion of the IEX. Roman numerals mark the different stages of the temporal evolution. I: Initial melting; II advance of the interface; III stationary interface; V degradation. An additional relaxation stage (IV) seems to be visible in the data of the sample with smaller concentration. But this stage in general was found much less clearly distinguishable and reproducible.

Fig. 7: Reduced interface position $y(\tilde{t})$ versus reduced time $\tilde{t}$ for various time-dependent drift velocities as obtained by a numerical solution of the nonlinear differential equation (7) for (a) y(0) = 0 and $g(\tilde{t}) = \Theta(\tilde{t})/2$; (b) y(0) = ln 2 after releasing g from g = 1/2 to zero as (i) $g(\tilde{t}) = \exp(-\tilde{t})/2$ (broken line), (ii) $g(\tilde{t}) = (1-\tilde{t})/2$ for 0 < $\tilde{t}$ < 1 and zero elsewhere (dotted line), (iii) $g(\tilde{t}) = \Theta(-\tilde{t})/2$ (full line). Here $\Theta(\tilde{t})$ denotes the unit step function.

Fig. 8: different morphologies obtained by melting aunder different boundary conditions. a) swiss cheese morphology frequently< observed during initial melting by homogeneously released carbonate (image height ca. 8mm); b) polycrystals slowly advancing and melting in a shallow gradient resemble a beach of pebble stones. In an extended coexistence region, these melt inward. (image height ca. 4.6mm); polycrystals in a strong drift, approaching the crystal melt interface in a steep gradient. Before melting they transform to large wall crystals of like crystallographic orientation. The wall crystals melt with straight sharp boundaries, showing kinks at their grain boundaries (image height ca. 5mm); d) columnar crystals grown in the absence of a drift (both taps open) and slowly melting in a moderate gradient as the carbonate concentration increases and showing extended kinks (image height ca. 5mm).